\documentclass[twocolumn]{aastex63}

\usepackage[squaren,Gray]{SIunits}
\usepackage{todonotes}
\newcommand{\sink}{\mathrm{sink}}

\newcommand\nmhd{$\textsc{NIMHD}$}
\newcommand\imhd{$\textsc{IMHD}$}
\newcommand\hydro{$\textsc{HYDRO}$}
\newcommand\nmhds{$\textsc{NIMHD}$ }
\newcommand\imhds{$\textsc{IMHD}$ }
\newcommand\hydros{$\textsc{HYDRO}$ }

\received{??}
\revised{??}
\accepted{??}

\submitjournal{ApJL}

\shorttitle{Disk birth in star forming clumps}
\shortauthors{Lebreuilly et al.}

\begin{document}

\title{Protoplanetary disk birth in massive star forming clumps: the essential role of the magnetic field}

\correspondingauthor{Ugo Lebreuilly}
\email{ugo.lebreuilly@cea.fr}

\author[0000-0001-8060-1890]{Ugo Lebreuilly}
\affiliation{AIM, CEA, CNRS, Université Paris-Saclay, Université Paris Diderot, Sorbonne Paris Cité, F-91191 Gif-sur-Yvette, France}
\nocollaboration{1}

\author[0000-0002-0472-7202]{Patrick Hennebelle}

\affiliation{AIM, CEA, CNRS, Université Paris-Saclay, Université Paris Diderot, Sorbonne Paris Cité, F-91191 Gif-sur-Yvette, France}
\nocollaboration{1}

\author[0000-0002-2636-4377]{Tine Colman}

\affiliation{AIM, CEA, CNRS, Université Paris-Saclay, Université Paris Diderot, Sorbonne Paris Cité, F-91191 Gif-sur-Yvette, France}
\nocollaboration{1}

\author[0000-0003-2407-1025]{Benoît Commerçon}
\affiliation{Univ Lyon, Ens de Lyon, Univ Lyon1, CNRS, Centre de Recherche Astrophysique de Lyon UMR5574, F-69007, Lyon, France}
\nocollaboration{2}

\author[0000-0002-0560-3172]{Ralf Klessen}
\affiliation{Universität Heidelberg, Zentrum für Astronomie, Institut für Theoretische Astrophysik, Albert-Ueberle-Str. 2, 69120 Heidelberg, Germany}
\affiliation{Universität Heidelberg, Interdisziplinäres Zentrum für Wissenschaftliches Rechnen, INF 205, 69120, Heidelberg, Germany}
\nocollaboration{4}

\author[0000-0002-3801-8754]{Anaëlle Maury}
\affiliation{AIM, CEA, CNRS, Université Paris-Saclay, Université Paris Diderot, Sorbonne Paris Cité, F-91191 Gif-sur-Yvette, France}
\nocollaboration{1}

\author[0000-0002-9826-7525]{Sergio Molinari}
\affiliation{Istituto Nazionale di Astrofisica - IAPS, Via Fosso del Cavaliere 100, I-00133 Roma, Italy}
\nocollaboration{5}

\author[0000-0003-1859-3070]{Leonardo Testi}
\affiliation{AIM, CEA, CNRS, Université Paris-Saclay, Université Paris Diderot, Sorbonne Paris Cité, F-91191 Gif-sur-Yvette, France}
\affiliation{ESO/European   Southern   Observatory,   Karl-Schwarzschild-Strasse   2   D-85748   Garching   bei   München,   Germany   e-mail:ltesti@eso.org}

\nocollaboration{3}

\begin{abstract}

Protoplanetary disks form through angular momentum conservation in collapsing dense cores. In this work, we perform the first simulations with a maximal resolution down to the astronomical unit (au) of protoplanetary disk formation, through the collapse of 1000~$M_{\odot}$ clumps, treating self-consistently both non-ideal magnetohydrodynamics with ambipolar diffusion as well as radiative transfer in the flux-limited diffusion approximation including stellar feedback. Using the adaptive mesh-refinement code {\ttfamily RAMSES}, we investigate the influence of the magnetic field on the disks properties with three models. We show that, without magnetic fields, a population dominated by large disks is formed, which is not consistent with Class~0 disk properties as estimated from observations.
The inclusion of magnetic field leads, through magnetic braking, to a very different evolution. When it is included, small $< 50~$au disks represent about half the population. In addition, about $\sim 70\%$ of the stars have no disk in this case which suggests that our resolution is still insufficient to preserve the smaller disks. With ambipolar diffusion, the proportion of small disks is also prominent and we report a flat mass distribution around $0.01-0.1 M_{\odot}$ and a typical disk-to-star mass ratios of $\sim 10^{-2}-10^{-1}$.
This work shows that the magnetic field and its evolution plays a prominent role in setting the initial properties of disk populations.
\end{abstract}


\section{Introduction}

Protoplanetary disks are a natural consequence of angular momentum conservation during the protostellar collapse. Planet formation in these disks not only depends on local quantities (density and temperature profile) but also on their global physical properties such as their total gas and dust masses or their size \citep[see][for a review on planet formation]{2014prpl.conf..339T}. Evolved disks, around Class~II-III young-stellar objects (YSOs), are now observed in significant numbers at high resolution \citep[see for example][]{2018ApJ...869L..41A}. They have typical sizes of $~100~$au \citep{2009ApJ...700.1502A,2021arXiv210111307S} but seem to be lacking the material to form giant planets \citep{2018A&A...618L...3M}, which suggests that those form early on. Unfortunately, the initial properties of protoplanetary disks are still poorly constrained. In contrast to Class~II-III disks, Young  Class~0-I disks are typically more compact and evolve over short lifetimes so they are difficult to observe. In addition, they are still deeply embedded in a massive envelope \citep{2002EAS.....3....1A} making their observation even more difficult. Following pioneering works, such as by \cite{2000ApJ...529..477L}, who inferred the presence of a disk around Class~0 objects, recent surveys such as CALYPSO \citep{2019A&A...621A..76M} and VANDAM \citep{2018ApJ...866..161S,2020ApJ...890..130T} have started to probe them more extensively. Although the disk component is often unresolved or marginally resolved at these stages, \cite{2019A&A...621A..76M} have reported typical radii of about $< 50~$au. Smaller disks could remain undetected because of insufficient resolution and studies such as by \cite{2015ApJ...812..129Y} even argued that objects, such as the Class~0 protostar B335, could have a disk smaller than $10$~au or even no disk at all.

On the theoretical side, our understanding of disk formation have gained in maturity over the two last decades. Historically, there has been an angular momentum problem in star formation. We know from observations, that the specific angular momentum of dense cores is not conserved during star formation and must be redistributed by an efficient physical process. Magnetic field, one of the most promising candidates, has been extensively investigated as a possible solution, both using ideal
\citep{2003ApJ...599..351A,2008A&A...477...25H,2012A&A...543A.128J}
and non-ideal \citep{2014prpl.conf..173L,2016A&A...587A..32M,2016MNRAS.463.4246M,2016MNRAS.460.2050Z,2018A&A...615A...5V,2016MNRAS.457.1037W,2020A&A...635A..67H,2020MNRAS.495.3795W,2021arXiv210207963L} 
magnetohydrodynamics (MHD). In both the ideal and non-ideal case, it was shown that magnetic braking would prevent the formation of large, massive and unrealistic disks such as those observed in purely hydrodynamical simulations \citep[see for example][]{2011MNRAS.413.2767M}. In fact, in ideal MHD it would in some cases, totally prevent disk formation. Several studies \cite[e.g.][]{2010A&A...521L..56D,2015ApJ...801..117T,2016ApJ...830L...8H} have shown that this so-called magnetic braking catastrophe could be solved by taking into account the role of diffusive processes, such as ambipolar diffusion, which reduces the braking efficiency at disk-like densities. In this case, a small disk with an initial size of $\sim 20~$au, that would grow later on, is expected. It was also shown that other effects such as the magnetic field misalignment \citep{2012A&A...543A.128J,2018MNRAS.473.2124G} or turbulence \citep{2012ApJ...747...21S} could also reduce the efficiency of the magnetic braking and lead to the formation of small disks.

In past studies, magnetized simulations have only been performed to study star formation in the case of low mass and isolated dense cores. In reality, these cores are connected to the large scale environment inside molecular clouds and most stars (and disks) are not born in isolation, but rather within turbulent and magnetized complexes with masses of $10^2-10^3 M_{\odot}$ called clumps \citep[see][for studies of massive clumps at the galactic scale]{2014MNRAS.443.1555U,2017MNRAS.471..100E}. To statistically understand protoplanetary disk formation, and in the light of the recent progress in observing very young disks populations, it is fundamental to model the protostellar collapse starting from massive clumps all the way down to the disks scale. To this date, only \cite{2018MNRAS.475.5618B} investigated disk formation in such clouds, although without including a magnetic field. Given the theoretical importance of the magnetic field in the isolated case, it is crucial to investigate, as well,  its impact within massive clumps.  

In this work, we therefore present a study of the disk populations resulting from the collapse of magnetized $1000 M_{\odot}$ clouds, in both the ideal and non-ideal MHD (with ambipolar diffusion) framework and also taking into account the radiative stellar feedback. We present three collapse calculations with a maximum resolution of $1.2~$au performed with the adaptive mesh-refinement (AMR) non-ideal MHD and radiative transfer code {\ttfamily RAMSES}. In section~\ref{sec:methods}, we describe our numerical methods. Then, in section~\ref{sec:results} we report on the disk populations obtained from our three models. Finally, we summarize our main results in section~\ref{sec:discussion}.

\section{Methods}
\label{sec:methods}

Our models are computed using the AMR \citep{1984JCoPh..53..484B} finite-volume code {\ttfamily RAMSES} \citep{2002A&A...385..337T,2006A&A...457..371F} and its extension to radiative transfer in the flux-limited diffusion (FLD) approximation \citep{2011A&A...529A..35C,2014A&A...563A..11C}, non-ideal MHD \citep{2012ApJS..201...24M} and sink particles \citep{2014MNRAS.445.4015B}.

Initially, we consider $1000 M_{\odot}$ uniform clumps of temperature $10~\kelvin$ and radius set according to the thermal-to-gravitational energy ratio\ $\alpha$ 
\begin{eqnarray}
\alpha \equiv \frac{5}{2} \frac{R_0 k_{\rm{B}} T_{0}}{\mathcal{G} M_0\mu_{\rm{g}}m_{\rm{H}}},
\end{eqnarray}
with $k_{\rm{B}}$ being the Boltzmann constant, $\mathcal{G}$ being the gravitational constant, $m_{\rm{H}}$ being the Hydrogen mass and $\mu_{\rm{g}}=2.31$ being the mean molecular weight. We set $\alpha=0.008$ and have an initial radius of $\sim 0.38$~pc and density of $3 \times 10^{-19}~\gram ~ \centi\meter^{-3}$. This initial condition is fairly typical of the clumps observed in the Milky-Way according to both the ATLASGAL \citep{2014MNRAS.443.1555U} and HI-GAL \citep{2017MNRAS.471..100E} surveys. We note that we also set a gas adiabatic index $\gamma=5/3$.

We set an initial turbulent velocity with a power-spectrum of $k^{-11/3}$,  which corresponds to a Kolmogorov spectrum, and random phases. The root-mean square of this velocity field is such that the initial Mach number is equal to 7 which corresponds to a  turbulent-to-gravitational energy ratio of about 0.4.

In two models, we set a vertical and uniform initial magnetic field using the mass-over-flux to critical-mass-over-flux ratio $\mu=10$ such as
\begin{equation}
    \mu = \left(\frac{M_0}{\phi}\right)/\left(\frac{M}{\phi}\right)_c,
\end{equation}
where $\left(\frac{M}{\phi}\right)_c = \frac{0.53}{\pi}\sqrt{5/\mathcal{G}}$ \citep{1976ApJ...210..326M}.  This corresponds to a magnetic field strength of $\sim 9.4 \times 10^{-5}$~G. In the non-ideal case, we only consider the effect of ambipolar diffusion which is most probably the dominant non-ideal mechanism at the density range that we consider \citep{2016A&A...592A..18M}, although  uncertainties remains about the strength of the Hall effect. The value of the ambipolar resistivity is computed as a function of the temperature, density and magnetic field intensity according to the table of \cite{2016A&A...592A..18M}. Similarly, the Planck and Rosseland opacities, used for the radiative transfer are computed using the tables described by \cite{2013A&A...557A..90V}.

To accurately follow the multiple scales of the clump, we use the AMR capability of {\ttfamily RAMSES}. The size of cell $\Delta x$ is given by 
$\Delta x = \frac{L_{\mathrm{box}}}{2^\ell}$
where $\ell$ is the level of refinement. In our models, we consider a $\sim 1.53$~pc box and an initially uniform grid with $\ell_{\mathrm{min}}= 7$ (corresponding to a $2460~$au or $0.012~$pc resolution) and we then refine the grid up to a maximum level $\ell_{\mathrm{max}}= 18$ ($1.2~$au or $5.84 \times 10^{-6}~$pc) according to the Jeans length $\lambda_{\mathrm{Jeans}}$ to impose
\begin{eqnarray}
\Delta x \leq \frac{\lambda_{\mathrm{Jeans}}}{N},
\end{eqnarray}
where $N>10$ to respect the \cite{1997ApJ...489L.179T} criterion and avoid artificial fragmentation.

In this work, we use sink particles \citep{2014MNRAS.445.4015B} to mimic the behavior of fully formed stars and avoid the numerical difficulties of increasing the numerical resolution enough to resolve them.  We form sink particles when the density reaches $n_{\rm{thre}}= 3 \times 10 ^{13}~ \centi\meter^{-3}$ as in the standard case of \cite{2020A&A...635A..67H}. Once a sink forms, it is placed at the position of the peak of the corresponding clump of density threshold $n_{\rm{thre}}/10$ and the gas within a region of $4 \Delta x$ (where $ \Delta x$ is the cell size) is accreted if its density is above $n_{\rm{thre}}/3$. After that, a  fraction $C_{\rm{acc}}=0.1$ of the mass above $n_{\rm{thre}}/3$ and within the accretion volume is attributed to the sink at each timestep. This corresponds to the fiducial value explored in \cite{2020A&A...635A..67H}. As they show, the value of this parameter does impact the evolution of the disk and should therefore be explored in the future.

A star, i.e. a sink, of mass $M_{\sink}$ and radius $R_{\star}$ is a source of luminosity when accreting mass. This so-called accretion luminosity is expressed as
\begin{eqnarray}
L_{\rm{acc}} = f_{\rm{acc}} \frac{\mathcal{G} M_{\sink} \dot{M}_{\sink}}{R_{\star}},
\end{eqnarray}
where $f_{\rm{acc}}\leq 1$ is the fraction of the accretion gravitational energy that is radiated away. In all our models, we consider $f_{\rm{acc}}=0.1$ which corresponds to the low value investigated by \cite{2020ApJ...904..194H}. In their study, they also investigated other values but concluded that for a cloud similar to the ones considered here, the value of this parameter did not affect much the stellar mass spectrum. The star radius and its luminosity $L_{\mathrm{int}}$, are computed using the models of \cite{2013ApJ...772...61K}.  Once the luminosities are computed, as in \cite{2020ApJ...904..194H} the corresponding energy $(L_{\mathrm{int}} +  L_{\rm{acc}} )\mathrm{d}t$ is uniformly distributed to the sink cloud particles over the sink accretion volume \citep[for more details on the cloud-in cell interpolation used see][]{2014MNRAS.445.4015B}.

\section{Results} 
\label{sec:results}
\begin{deluxetable*}{ccccccccc}
\tablenum{1}
\tablecaption{\label{tab:table1}
Summary of the three models. From left to right: name of the model, number of stars (i.e. sinks), number of isolated/highly-separated stars, number of disks and close multiple systems,  $t|_{50 M_{\odot}}$ and $t|_{160 M_{\odot}}$, median disk mass and median disk radius. All the quantities with a bar are time averaged between $t|_{50 M_{\odot}}$ and  $t|_{160 M_{\odot}}$, the other quantities are measured at $t|_{160 M_{\odot}}$.} 
\tablehead{Name &   $N_{\rm{star}}$ &  $N_{\mathrm{isol}}$&  $N_{\mathrm{disks}}$ & $N_{\mathrm{syst}}$  & $t|_{50 M_{\odot}}$ [kyr]& $t|_{160 M_{\odot}}$ [kyr] & $\overline{M}_{\mathrm{disk}}$ [$M_{\odot}$] & $\overline{R}_{\mathrm{disk}}$~[au]}
\startdata
\imhd  &  147 & 73  & 31 & 16 & 105  & 118  & 0.021 & 50.6 \\
\nmhd & 191 & 104 & 42 & 18  & 103 & 117 & 0.041 & 46.3 \\
\hydro & 212 &  128 &102 & 27 &  98 & 112 & 0.037 & 60   \\
\enddata
\end{deluxetable*}
\begin{figure*}[t!]
    \includegraphics[width=
         \textwidth]{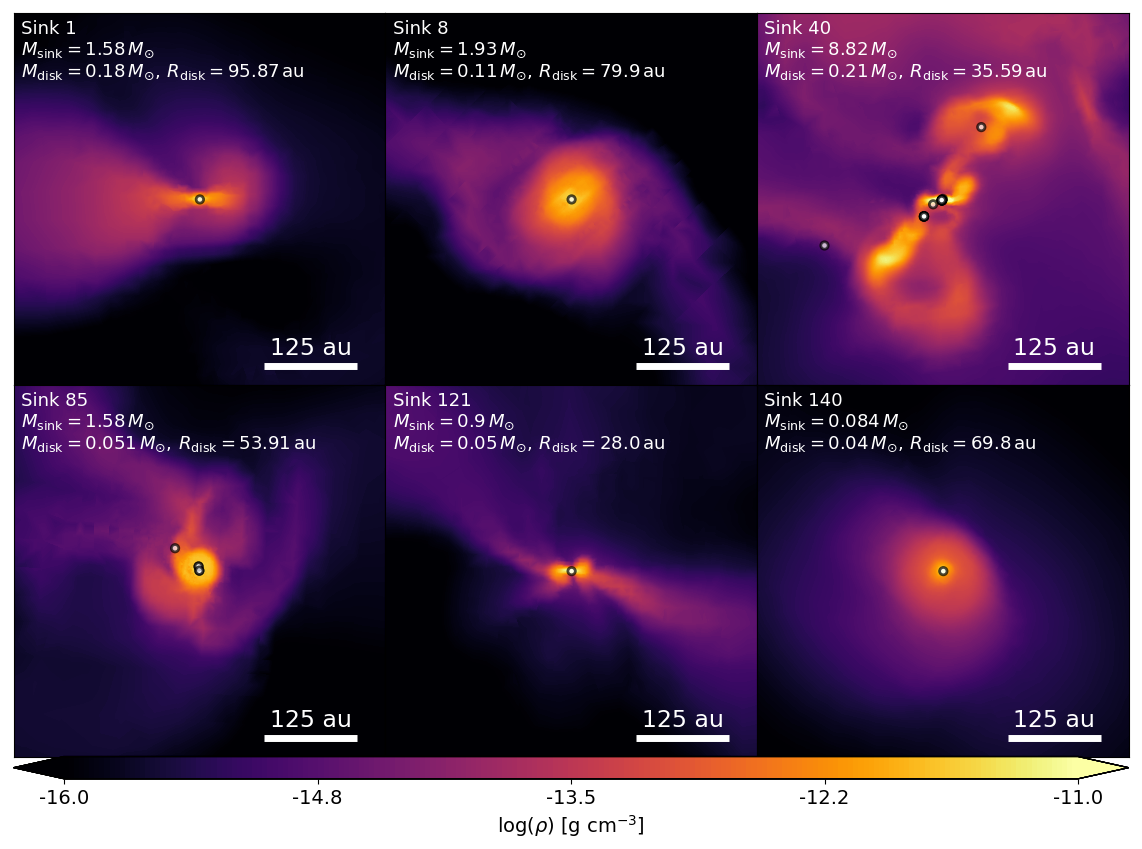}
         \includegraphics[width=
         \textwidth]{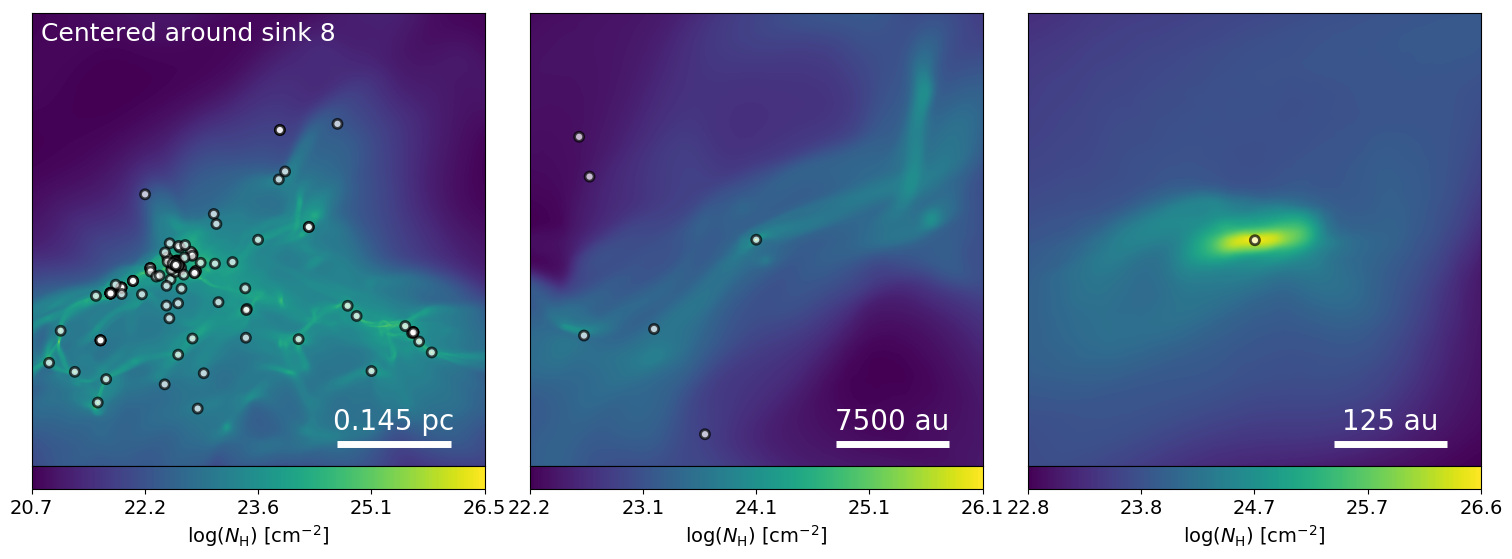}
\caption{\nmhds model at $t|_{160 M_{\odot}}=117~$kyr. (Top 6) Density slices for various disks around primary stars. The colorbar and spatial extension of the six slices are the same for the sake of comparison and they are alternatively displayed edge-on or face-on. Finally, we display the sink number, the stellar and disk mass and the disk radius for each density snapshot. (Bottom 3) Edge-on column density slices at various scales centered around sink 8. Circles represent sinks for all the snapshots and all of them are centered around the primary.} 
\label{fig:disknmhd}
\end{figure*}

\begin{figure*}[t!]
\centering
          \includegraphics[width=
         \textwidth]{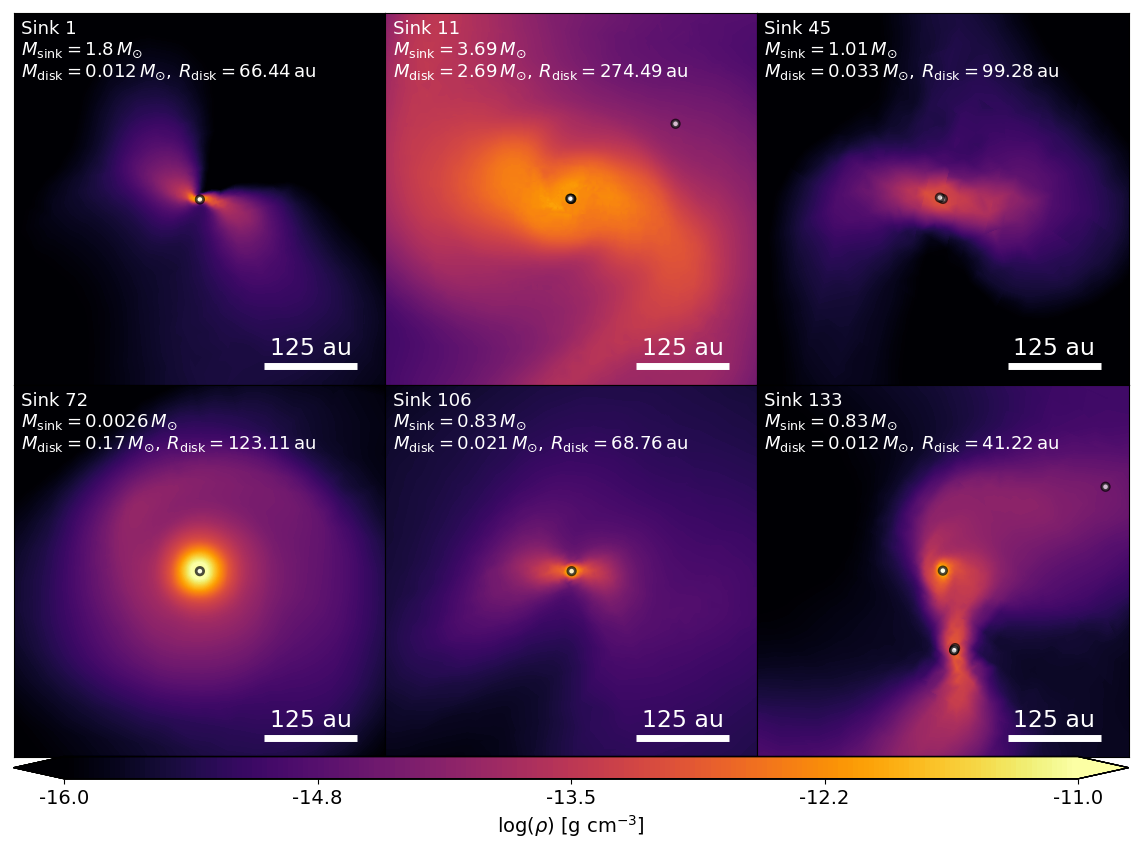}
      \caption{ \imhds at $t|_{160 M_{\odot}}=118~$kyr. Same as the top six snapshots of figure \ref{fig:disknmhd}.} 
            \label{fig:diskimhd}
\end{figure*}
\begin{figure*}[t!]
\centering
          \includegraphics[width=
         \textwidth]{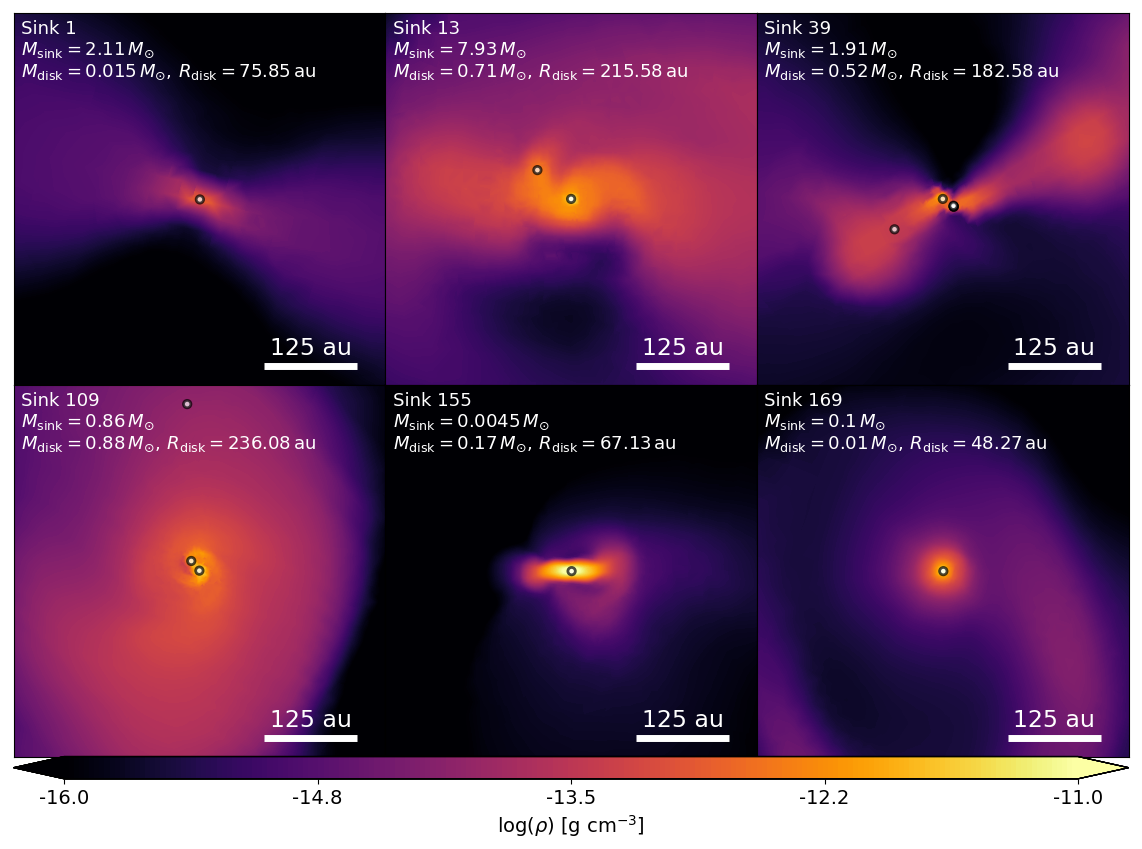}
      \caption{\hydros at $t|_{160 M_{\odot}}=112~$kyr.  Same as the top six snapshots of figure \ref{fig:disknmhd}.} 
      \label{fig:diskhydro}
\end{figure*}
\begin{figure*}
\centering
          \includegraphics[width=0.8\textwidth]{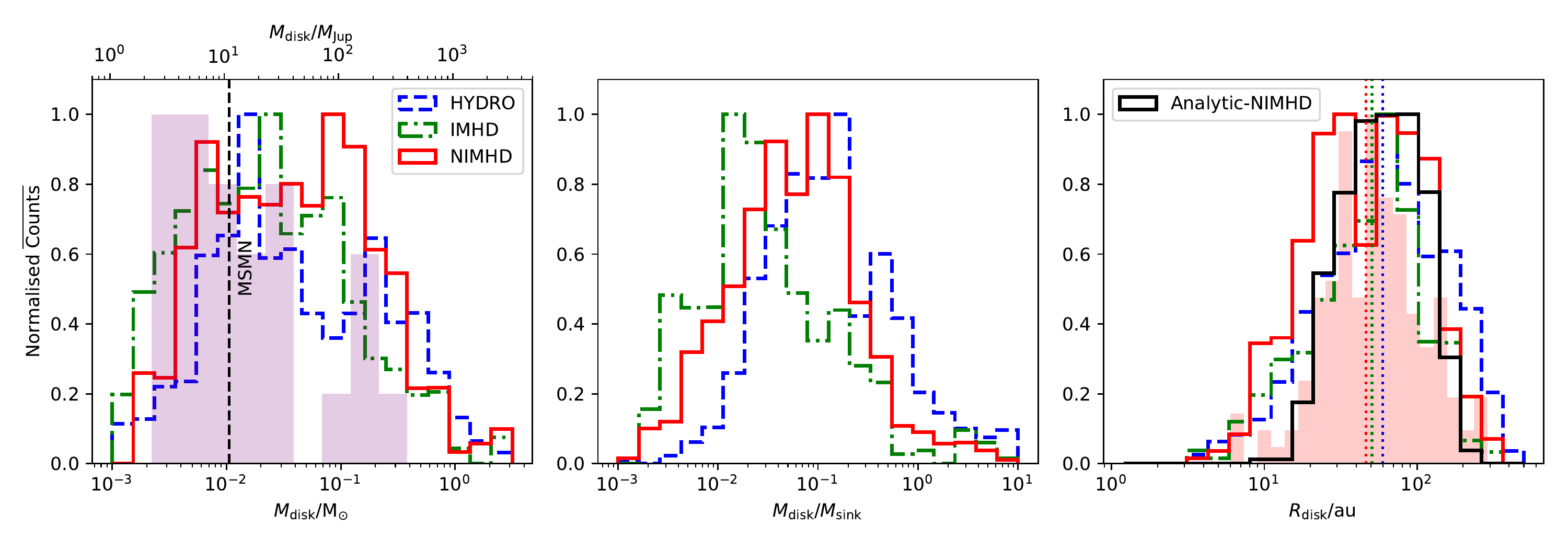}
          \includegraphics[width=0.8\textwidth]{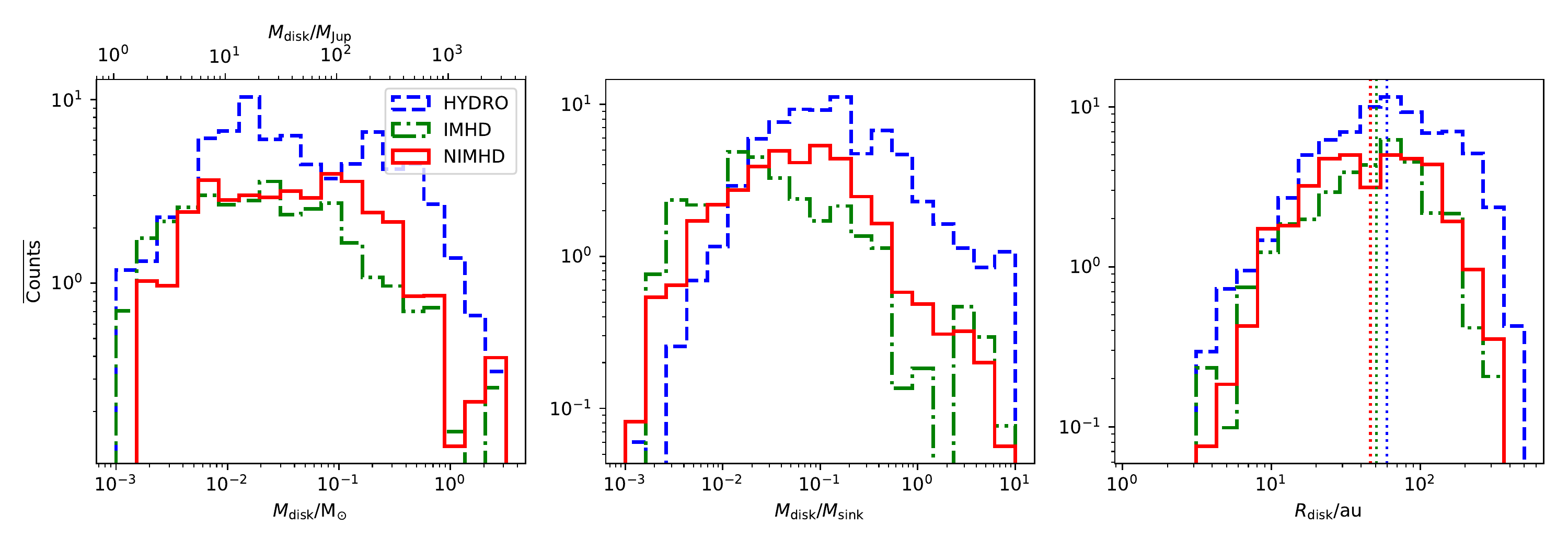}
        \includegraphics[width=0.8\textwidth]{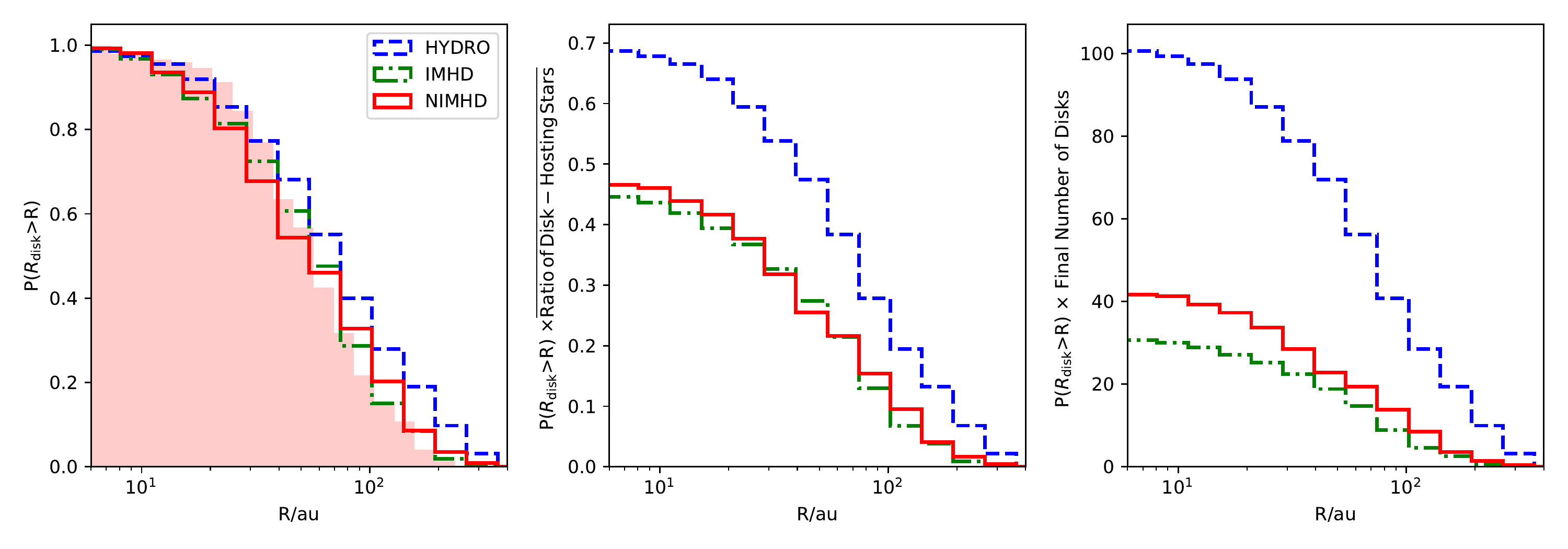}
      \caption{ (Top) Distributions (normalized so that their peak in the y-axis is 1) for the three models. (Left) Mass of the disks in solar mass (bottom x-axis) and Jupiter mass (top x-axis), minimum solar mass nebula  \citep{1981PThPS..70...35H} (dashed line), VLA Class~0 disk mass distribution (purple) inferred by \cite{2020A&A...640A..19T}. (Center) Disk to primary mass ratio. (Right) Disk radius distribution, medians radii of the models (dashed vertical lines, same color as the models), analytical estimate \citep{2016ApJ...830L...8H} (black line), Class~0  disks (pink) from the merged CALYPSO \citep[][]{2019A&A...621A..76M} and VANDAM \citep{2018ApJ...866..161S,2020ApJ...890..130T} surveys.
      (Middle) Same but with no normalization of the histograms and without the observations. 
      (Bottom) Cumulative distribution function of disk radius for the three models. Left: normalized. Middle: normalized  and multiplied by the ratio disk-hosting stars. Right: normalized and multiplied by the number of disks. 
      All distribution are time averaged (between $t|_{50 M_{\odot}}$ and  $t|_{160 M_{\odot}}$) distributions. }
            \label{fig:hist}
\end{figure*}

We now introduce our three models. First, two magnetized models were computed with (\nmhd) and without (\imhd) ambipolar diffusion. The third model (\hydro), is similar but without any magnetic field. The main properties of the models are summarized in table~\ref{tab:table1}. 
We integrated them until the cumulative mass in sink particles is $160~M_{\odot}$, which is $16\%$ of the initial cloud mass. For the sake of comparison, averages are made between $t|_{50 M_{\odot}}$ and  $t|_{160 M_{\odot}}$ where $50 M_{\odot}$ and $160 M_{\odot}$ have been accreted, respectively.

We define the number of isolated stars as the number of stars without any close neighbor. It implicitly includes stars that are in long-period multiple systems, the study of which  is beyond the scope of this paper. Our definition of a close system indeed only considers stars within less than $50~$au (see appendix). We also define the number of systems as the number of primary stars that have at least one close neighbor\footnote{by definition $N_{\rm{star}} \ne N_{\rm{isol}}+N_{\rm{syst}}$}.
 
At the end of the calculation, \nmhds has 191 sinks/stars, 18 close multiple systems, 104 sinks in isolation and 42 disks are detected. \hydros has 212 sinks, which includes 128 sinks in isolation, around which we find 102 disks. Finally, \imhds only has 147 sinks, including 73 in isolation and we detect only 31 disks. The varying number of sinks in the three models suggests that fragmentation is significantly suppressed in \imhds but is also mostly recovered in \nmhd.

The mean disk masses of the three models are comparable although the value is slightly lower for \imhds for which it is $ \sim 0.021 M_{\odot}$, against $ \sim 0.041 M_{\odot}$ for \nmhds and $0.037  M_{\odot}$ for \hydro.  In terms of radii, \hydros has the largest disks with a median radius of $\sim 60~$au. Both \imhds and \nmhds produce an extra population of small disks because of magnetic braking. For that reason the median disk radius is smaller for the two magnetized models ($\sim 46.26$~au for \nmhds and $\sim 50.6$~au for \imhd). Besides reducing the typical size of the disks, the magnetic braking is in fact very damaging for the population of small disks. As previously mentioned, the \imhds model only have 31 disks at the end of the simulation while \nmhds has 42. The ratio of disk-hosting stars is lower by a factor of almost two in the magnetized models. This suggests that, although more disks form through the course of the simulation, the magnetic braking strongly affects the very small disks ($< 10-15~$au) to the point that some of them essentially become unresolved and disrupted at the grid scale. The smaller disks of the magnetized runs indeed typically have their rotation axis aligned with the large-scale magnetic field which maximizes the braking efficiency \citep[see][for studies of the impact of misalignment on disks properties]{2012A&A...543A.128J,2018MNRAS.473.2124G}. This critical effect should be investigated in future works with various initial field strengths. The values presented above are summarized in table~\ref{tab:table1}.

Figures \ref{fig:disknmhd}, \ref{fig:diskimhd} and \ref{fig:diskhydro} show density slices for the \nmhd, \imhds and \hydros models, respectively. The disks are displayed at the end of the simulation, when $160 M_{\odot}$, from the initial $1000 M_{\odot}$, have been transformed into stars. In figure \ref{fig:disknmhd}, and more generally in the \nmhds model, the disks typically have small radii, among the six ones presented here (disks 1, 8, 40, 85, 121 and 140), half of them are smaller that $55~$au and the others are smaller than $100$~au. Their size is most likely controlled by magnetic self-regulation that favors the formation of small disks \citep{2016ApJ...830L...8H}. The disks of \hydros and \imhds are on average larger, although for very different reasons. In the case of \imhd, this is only a slight difference and small disks are still often formed by the effect of the magnetic braking, but they are typically short-lived as no diffusive effect is counter-balancing the braking. Eventually, the population of large, hydro-like, disks survives better. For \hydro, small disks can form, but they are statistically rarer as no efficient mechanism leads to angular momentum transport outside the disks. The 12 disks displayed in the respective panels of the two models (figure \ref{fig:diskimhd} and \ref{fig:diskhydro}) are again fairly typical and have radii of the order of $80-100~$au (and up to $\sim 200~$au for the largest ones presented here).

In figure \ref{fig:disknmhd}, we added (bottom 3 pictures) edge-on column density slices centered around sink 8 at various scales ($0.58 ~$pc, $3 \times 10^4~$au and $500~$au). From the bottom-left panel, it appears quite clear that stars are not evenly spread in the cloud but rather concentrated within a cluster of about $1-2 \times 10^4~$au. As can be seen (top-middle panel and bottom-right panel), sink/disk 8 is not deeply embedded in this cluster and is actually quite isolated. It is therefore unsurprising that its disk, but also the ones around sink 121 and 140 (that are also quite isolated), appear to be mostly unperturbed and  well organized. Conversely, some stars in the cluster (such as sink 40) are quite significantly perturbed by close interactions. Generally speaking, interactions between the disks and neighboring stars are significant for the three models. The example of sink 40 in \nmhds model is striking, as it is part of the aforementioned cluster and highly perturbed by the other stars born in the same filamentary structure. Similar interactions and flybys are also happening in \imhds and \hydros (e.g. disk 133 for \imhds or disk 13, 39 and 109 for \hydro). Among the three models, \hydros shows more of these interactions for mainly two reasons; i) because disks are larger and are therefore affected by flybys over larger distances ii) because the \hydros disks are fragmenting more efficiently. 

All three models show a wide diversity of sub-structures and commonly observed \citep[around more evolved YSOs][]{2020arXiv200105007A} disk features such as spirals (often associated with the presence of a companion e.g. disk 11 and 72 in \imhd), warps (e.g. disk 1 in \imhd), accretion streamers (e.g. disk 121 in \nmhd) and circum-multiplicity (e.g. disk 109 in \hydro). We also note that disks tend
to be either strongly perturbed or relatively quiet and isolated (e.g. disk 155 in \hydro).
In the former case this is either because of star-disk interaction or because disks accrete high density material (e.g. disk 85 in \nmhds which is also a circumbinary disk according to our definition, as are disk 45 in \imhds and disks 39 and 109 in \hydro). We note that we do not see clear outflows around most stars in our models, which is likely a consequence of i) the high level turbulence that reduces the 
coherence of the flow and ii) the lack of resolution which precludes the launching of disk winds and jets.

 In figure~\ref{fig:hist} top panels, we show the distributions, normalized to have a y-axis that peaks at 1 (to compare their shapes), of the disk masses (left), the disk-to-star mass ratio (center) and the disk radius (right) for the three models (denoted by the three colors). We add a vertical line that represents the minimum solar mass nebular \citep[MSMN][]{1981PThPS..70...35H}, the information in terms of Jupiter mass ($M_{\rm{Jup}}$) and the Class~0 disks mass as estimated by \cite{2020A&A...640A..19T} and converted into gas mass assuming dust-to-gas ratio of $1\%$ in the disk mass histogram. In the disk radius histogram we add the Class~0 disks distributions of the CALYPSO \citep[][and references therein]{2019A&A...621A..76M} and VANDAM  \citep{2018ApJ...866..161S,2020ApJ...890..130T} surveys merged together (in pink), we also add three vertical dotted lines that represent the median disk radius of the models. We stress that the initial conditions used in this work are likely too compact compared to the clouds observed by these two surveys. However, so far they are the only statistical samples available at the Class~0 stage and it is therefore worth to make this comparison keeping in mind the possible bias. Finally, we also recompute a disk radius distribution from the mass of our primaries using the analytical estimate $r_{\mathrm{ana}}$ from \cite{2016ApJ...830L...8H} that writes as 
\begin{eqnarray}
r_{\mathrm{ana}} =18~\mathrm{au}~\delta^{2/9} & &  \left(
\frac{\eta_{\mathrm{AD}}}{0.1 \mathrm{s}}\right)^{2/9} \left(\frac{B_{z}}{0.1 \mathrm{G}}\right)^{-4/9} \nonumber \\ & &\left(\frac{M_{\mathrm{disk}}+M_{\sink}}{0.1 M_{\odot}}\right)^{1/3},
\end{eqnarray}
where $\delta$ is a coefficient of the order of a few, $\eta_{\mathrm{AD}}$ is the ambipolar resistivity, $B_{z}$ is the vertical magnetic field in the disk and $M_{\mathrm{disk}}$ and $M_{\sink}$ are the disk and star mass, respectively. For simplicity we assumed $\delta=1$ and $\eta_{\mathrm{AD}}=0.1 \mathrm{s}$ (as in the aforementioned study) while the other quantities are volume averaged within the disk. As a complementary information, the three middle panels of the figure show the non-normalized distributions that helps to compare the three models directly. The bottom panels show the cumulative distribution function of the disk radius with three different normalization (left: normalized as a probability, middle: normalized by the ratio of disk-hosting star, right: normalized by the number of disks).

 The three mass distributions are quite different, although their typical values are similar and range between $10^{-2}$ and $10^{-1} M_{\odot}$. As could be expected, \hydros forms the most massive disks while \imhds forms the less massive ones despite forming the most massive stars. In both magnetic runs, some large disks still form when the magnetic field is misaligned and the braking efficiency is low. For \imhd, particularly, they are in fact even more stable than their hydrodynamical counterpart because their rotation generate a strong toroidal magnetic field that stabilize them against fragmentation (as it is for example the case for disk 72 in figure~\ref{fig:diskimhd}). Contrary to the two other models, \nmhds has a flat disk mass distribution between $\sim 5 \times 10^{-2} M_{\odot}$ and $\sim 10^{-1} M_{\odot}$. Let us now concentrate on the disk-to-star mass ratio histogram. In \hydro, it peaks around $0.05-0.2$, which is very similar to what \cite{2018MNRAS.475.5618B} has found in a previous study. \imhds forms less massive disks relatively to their parent star because stars are typically more massive as the level of fragmentation of this model is reduced. Finally, the distribution of \nmhds lies in between the ones of \hydros and \imhds and peaks around $\sim 0.03-0.2$.
In all the cases, the disk-to-star mass ratio histograms are quite peaked, which shows that the mass of the star and the mass of the disk are correlated. A notable spread of about one order of magnitude indicates however that the stellar mass is not the only parameter that controls the disk mass.

We now focus on the disk radius histogram. We see that the \nmhds and \imhds models both form a population of small disks ($<50$~au) which represents about half of the disks. The \hydros distribution of the radius peaks close to $\sim 60-70~$au. \hydros presents the largest fraction of very large disks, with a prominent tail for the distributions between $100~$au and $300-400~$au, whereas the large disk population of both \nmhds and \imhds typically sharply decreases around $100-200~$au.  Although the shapes of the three radius histograms are quite similar, the three distributions are in fact very different as the three models do not form the same number of disks. This is shown clearly in the middle and bottom panels of figure~\ref{fig:hist} where we display non-normalised histograms (of the disk mass, the disk-to-star mass ratio and the disk radius) and cumulative distributions (of the disk radius) with three different normalization, respectively. Since about $55\%$ of the stars are in fact without a disk or with an unresolved disk component in \nmhds and \imhds. This means that only $\sim 20-25 \%$ of the stars have a disk larger than $50~$au in these runs. The existence of a disk around these stars is uncertain and should be studied at higher resolution, which is not possible with our current numerical capabilities. It is however strongly suggested by high resolution studies such as for example
by \citet{2010A&A...521L..56D,2018A&A...615A...5V}, that very small disks could form around some stars.
We note that the population of small disks is slightly larger in \nmhd, which is consistent with the regulation of magnetic braking by ambipolar diffusion. Larger statistical samples would however be useful to better access this difference.  Comparing the \nmhds PDF and the one obtained with the analytical model, we find a reasonable agreement between the two distributions although the analytical one presents a narrower range and has a less pronounced small disk population.

We superimposed the distributions of the radius of Class~0 disks extracted from the CALYPSO and VANDAM surveys merged together. Their histogram typically peaks around $50-60$~au and agrees best with the ones from the two magnetized models. We note that our tentative comparison relies on both the observational and numerical definitions of a disk that both need to be questioned. It also relies on the underlying models that are used to estimate the disks properties from observations \citep[for example assuming $1\%$ of dust which is not necessarily correct see for e.g.,][]{2020A&A...641A.112L} and on detection bias, as very small disks might be unresolved around Class~0 protostars. In addition, our clouds have different conditions as the ones probed by these surveys (e.g. in the Perseus and Orion region), we rather model more compact clouds that have more affordable simulation cost. Future comparisons with real disks should be dedicated to generate synthetic observations of the models in order to ensure that the observational and numerical definition of a disk are in agreement.  

\section{Conclusion}
\label{sec:discussion}

In this work, we presented the first collapse calculations of massive $1000 M_{\odot}$ clouds having au spatial resolution, which include a full treatment of the radiative transfer with stellar feedback as well as non-ideal MHD with ambipolar diffusion. Through three calculations, we explored the impact of the magnetic field on the disk populations self-consistently formed in the calculations.Our main findings and conclusions are:

\begin{itemize}
\item In the three models, we extract disk populations and infer their disk mass and radius distributions.
\item In the hydrodynamical case, and in accordance with the previous study of \cite{2018MNRAS.475.5618B}, we form massive and radially extended disks, often prone to fragmentation. We also note that, on average, about $70 \%$ of systems (single or multiple) have a resolved disk in this model.
\item In both the non-ideal and ideal MHD runs, we report a population of small disks. With ambipolar diffusion, the population is slightly prominent as magnetic braking is regulated by ambipolar diffusion. For both models, we note that the magnetic braking is particularly damaging for the population of small disks, leading to a low time-averaged ratio of disk-hosting stars about $44-45 \%$ in both models. As mentioned in section \ref{sec:results}, very small disks might still form around the remaining stars if those were not dissipated at the grid scale, i.e., with a higher resolution. This stresses the need for development of new numerical methods to meet the challenge of computing more resolved models.
\item The disks formed in our calculations are initially massive enough to host solar-like exoplanetary systems. More than half of the disks are more massive than the minimum solar mass nebula (MSMN) in all our models. This assertion is strengthened by the fact that the limit has been downward revised in more recent theories \citep{2007ApJ...671..878D}.
\item We produce a wide diversity of structures in our disks such as spirals, warps that are typical of older Class~II-III objects. We also frequently observe stellar encounters and close multiplicity. Future dedicated studies should investigate whether they are related or whether the ones around older objects come from a different origin.
\item Fragmentation is significantly reduced in the ideal MHD case but mostly retrieved when including ambipolar diffusion.
\end{itemize}

\acknowledgments
We thank the referee for providing useful comments that helped us to significantly improve our manuscript. This work is funded by the ERC synergy ECOGAL (Grant : 855130, PIs: P. Hennebelle R. Klessen,  S. Molinari, L. Testi) and was granted access to the HPC resources of CINES (OCCIGEN) under the allocation DARI A0090407023 made by GENCI. The figures were produced with matplotlib (\url{https://matplotlib.org/}) using the OSYRIS library (\url{https://pypi.org/project/osyris/}) develloped by Neil Vaytet whom we thank. The data presented in this article will be made available on the Galactica database (\url{https://galactica-simulations.eu/db/}) developed by Damien Chapon whom we also thank.

\appendix
\section{Disk finder}

\begin{figure*}
\centering
          \includegraphics[width=0.45\textwidth]{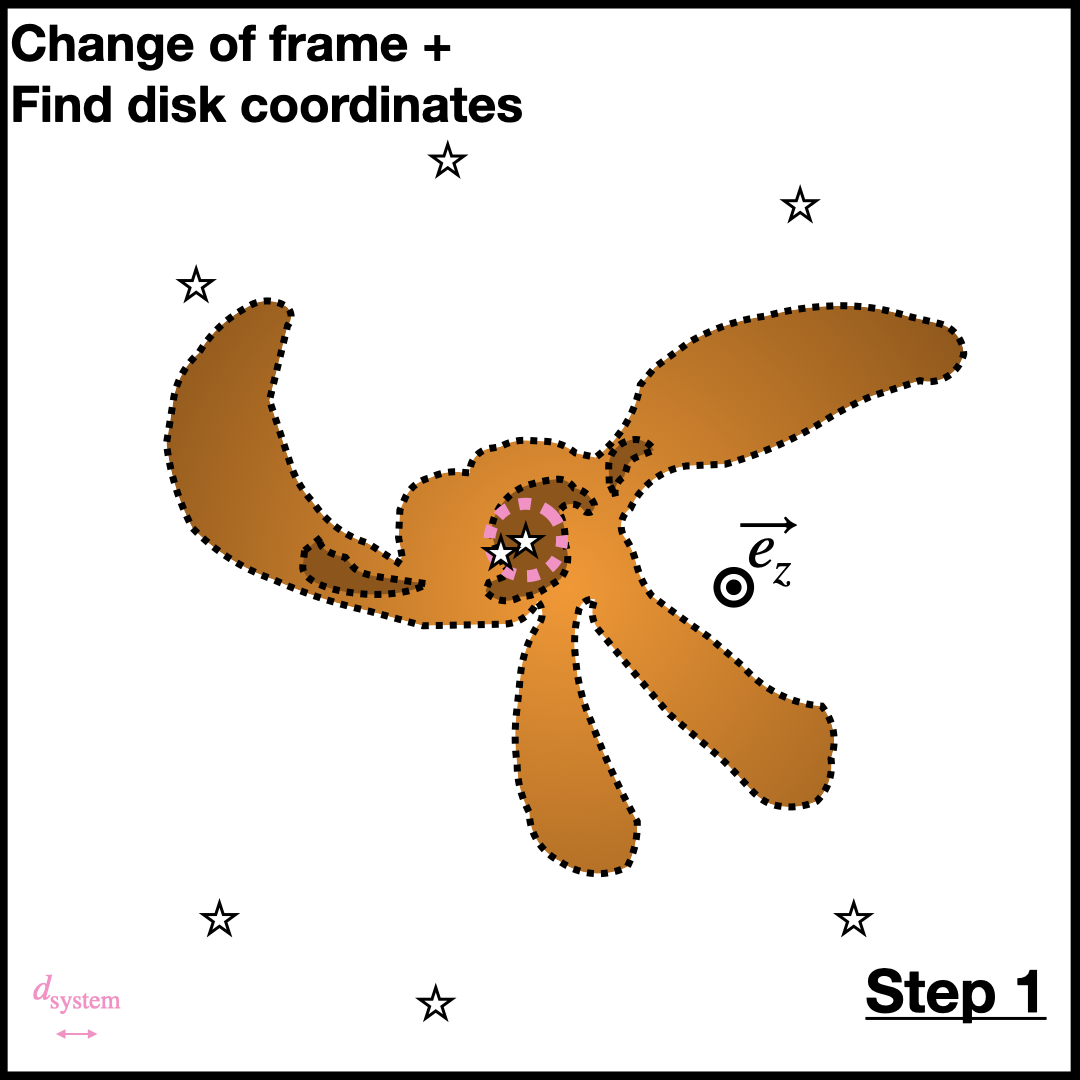}
        \includegraphics[width=0.45\textwidth]{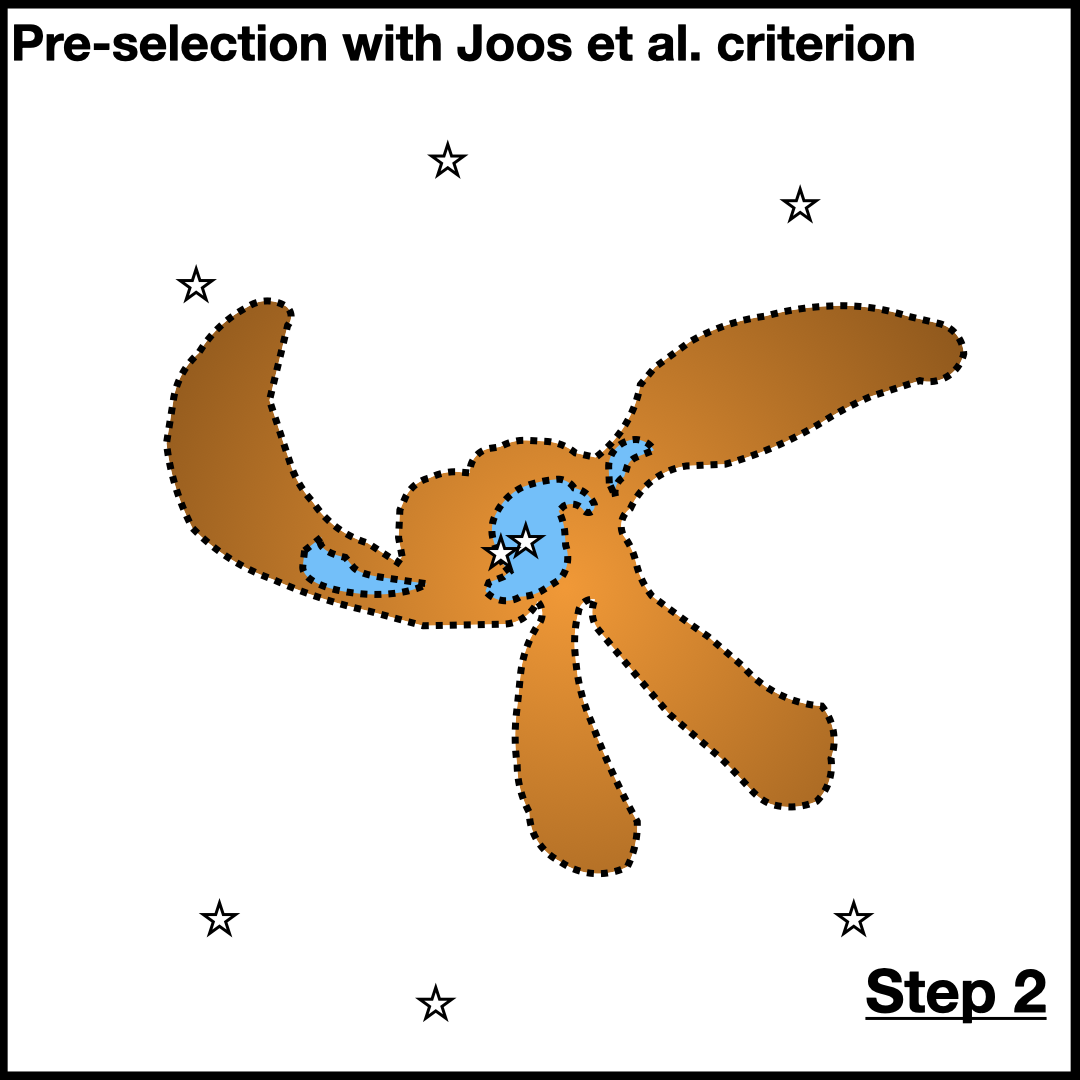}
\\ 
          \includegraphics[width=0.45\textwidth]{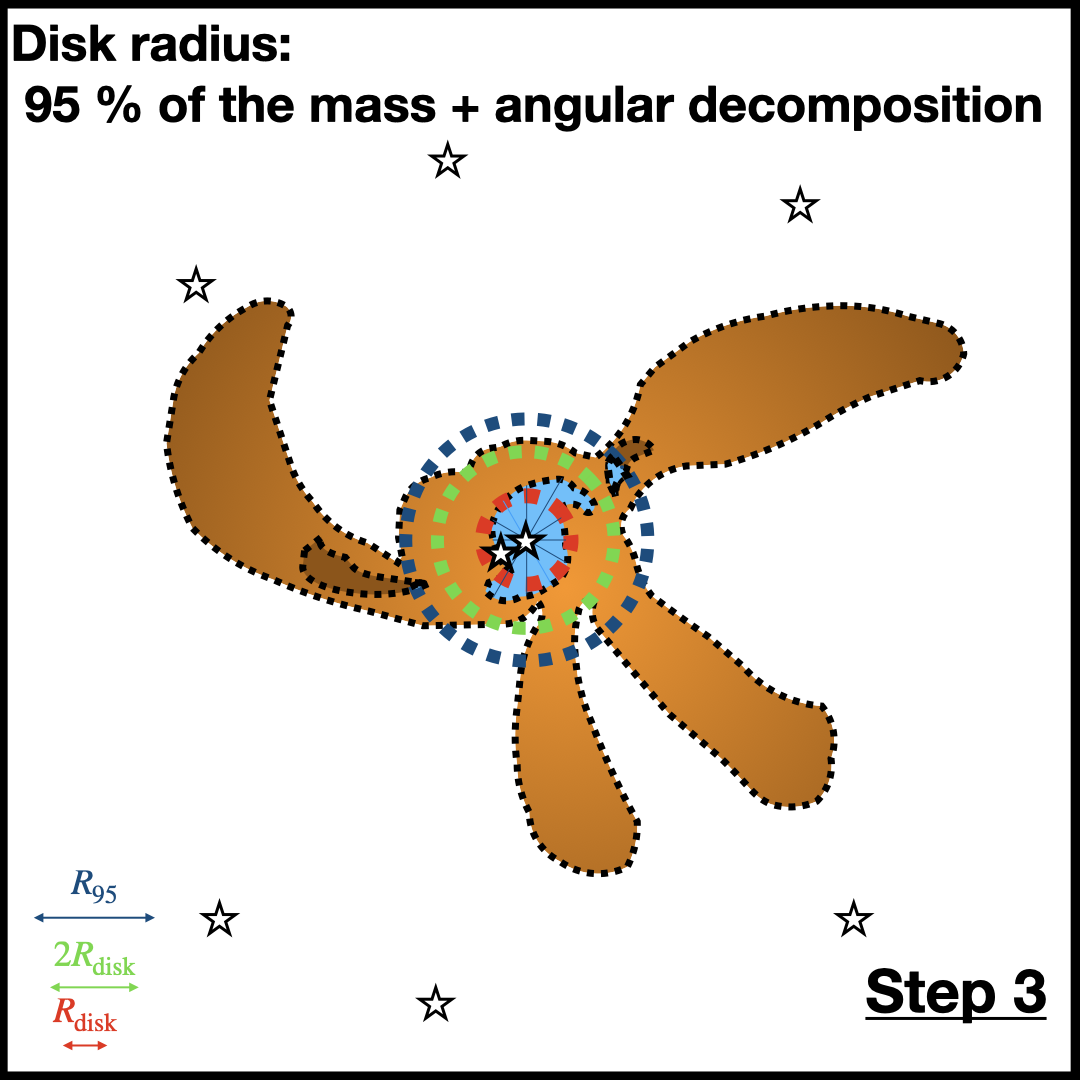}
          \includegraphics[width=0.45\textwidth]{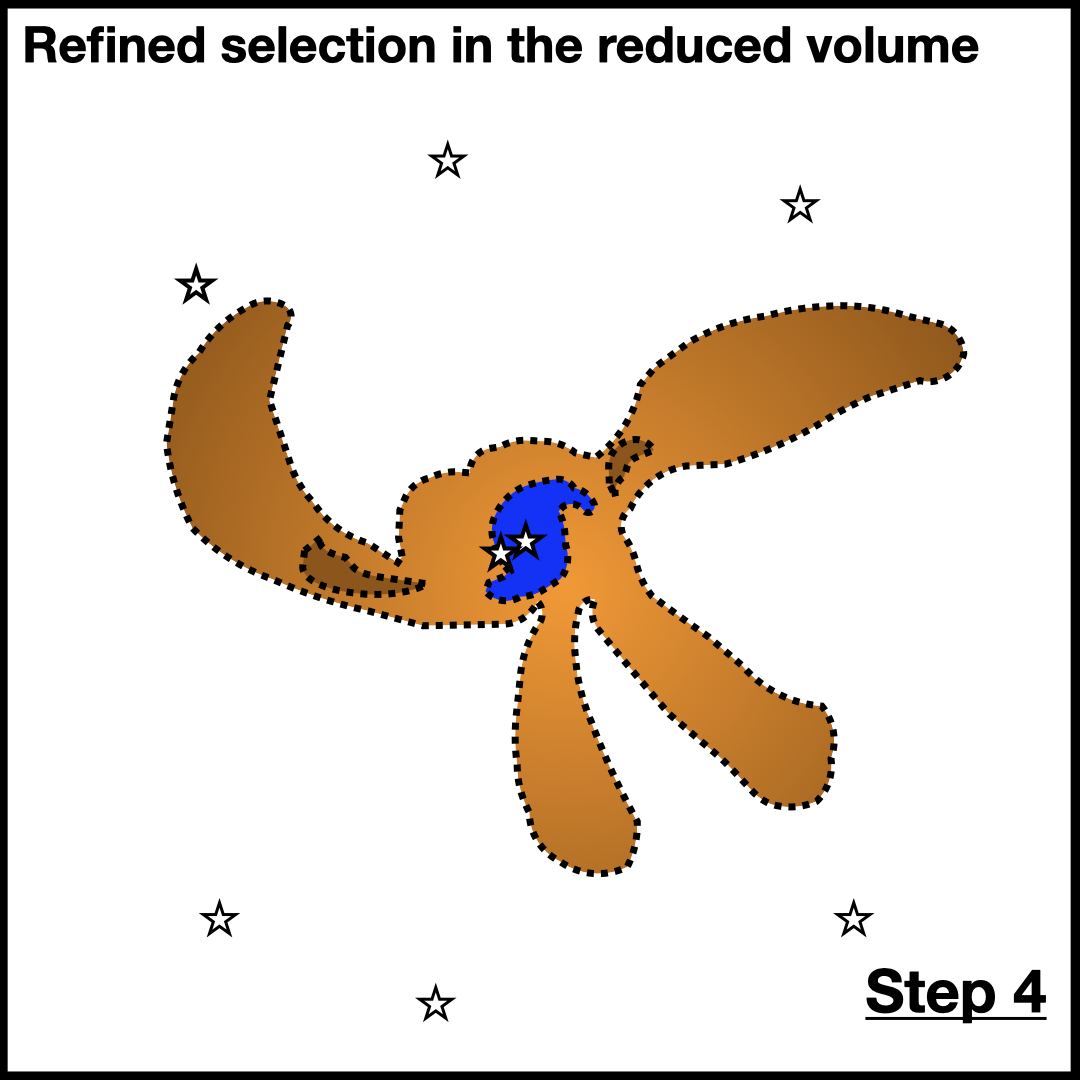}

      \caption{Schematic view of the four disk finder steps. i) change of frame and  coordinates ii) pre-selection iii) estimation of the radius iv) refined final selection.}
            \label{fig:disk_find}
\end{figure*}

 The disk selection process is a challenging task. Disks are indeed poorly defined and have arbitrary geometries. In addition they can easily be confused with dense free-falling filamentary material and stars are often born in multiple systems. In this work, we propose a new way to isolate at best the disk material which is largely inspired from \cite{2012A&A...543A.128J} who used the following criteria, 
 for a gas cell to be considered being included in the disk, it should:
\begin{itemize}
    \item rotate faster than it falls radially  $v_{\rm{\phi}} > 2 v_r$,
    \item rotate faster than it falls vertically $v_{\rm{\phi}} > 2 v_z$,
    \item not be thermally supported $\frac{1}{2} \rho v_{\rm{\phi}}^2 > 2 P_{\rm{th}}$, where $P_{\rm{th}}$ is the thermal pressure,
    \item be composed of dense material $n > n_{\rm{thre}} = 10^9~\centi\meter^{-3}$, where $n$ is the gas number density.
\end{itemize}

In our case, the disk selection operates as follows:
\begin{enumerate}
   \item We place ourselves in the co-moving frame of the barycenter of stars within a distance $r_{\mathrm{system}}=50~$au from the analyzed sink. Note that we only analyze disks around primary stars (the most massive star within $r<r_{\mathrm{system}}=50~$au).
   We then get the disk rotation axis by computing the direction of the angular momentum in a pre-selected region of radius $R_{\mathrm{mom}}= 15~$au.
    \item We select the cells that verify the \cite{2012A&A...543A.128J} criterion in the disk frame.
    \item We aim to exclude the streamers i.e., the large scale free-falling material. We first select the region within the cylindrical radius $R_{95}$ that encloses $95\%$ of the mass. We also do the same in the vertical direction. Then we decompose the disk in $N_{\theta}=50$ regions around the polar axis in the mid-plane (within $4\Delta x$) to compute the local radius. At this stage, we skip the analysis if there are no disk element in more than 2/3 of the regions. Otherwise, we compute the disk radius $R_{\rm{disk}}$ as the azimuthal median of the local radii.
    \item We re-apply the \cite{2012A&A...543A.128J} criterion in the regions where $r<R_{\mathrm{thre}}=\mathrm{min}(2~R_{\rm{disk}},R_{95})$. The arbitrary pre-factor of 2 allows to preserve some of the non-axisymmetric features (e.g., spirals). 
\end{enumerate}

In figure \ref{fig:disk_find}, we show a schematic simplified view of the four disk finder steps.

\bibliography{ref}
\bibliographystyle{aasjournal}

\end{document}